# Finding the best design parameters for optical nanostructures using reinforcement learning

Iman Sajedian[1], Trevon Badloe[1] and Junsuk Rho[1,2*]

[1] Department of Mechanical Engineering, Pohang University of Science and Technology (POSTECH), Pohang 37673, Republic of Korea

[2] Department of Chemical Engineering, Pohang University of Science and Technology (POSTECH), Pohang 37673, Republic of Korea

* Corresponding author email: jsrho@postech.ac.kr

**Abstract**

Recently, a novel machine learning model has emerged in the field of reinforcement learning known as deep Q-learning. This model is capable of finding the best possible solution in systems consisting of millions of choices, without ever experiencing it before, and has been used to beat the best human minds at complex games such as, Go and chess, which both have a huge number of possible decisions and outcomes for each move. With a human-level intelligence, it has been solved the problems that no other machine learning model could do before. Here, we show the steps needed for implementing this model on an optical problem. We investigated the colour generation by dielectric nanostructures and show that this model can find geometrical properties that can generate a much deeper red, green and blue colours compared to the ones found by human researchers. This technique can easily be extended to predict and find the best design parameters for other optical structures.

**Introduction**

Plasmonic structures can be used to create high resolution images beyond the diffraction limit of light[1-4]. Due to their small dimensions, the resolution can be as high as 100k dots per inch[5-7]. Unfortunately, they suffer from optical losses and poor colour saturation which leads to a small colour gamut[4]. Many designs using different types of shapes, structures and materials have been proposed to expand the achievable colour gamut as much as possible[8-11]. One design of note is based on an all dielectric nanostructure using silicon rods with an antireflective layer[8]. The colour gamut achieved was wider than previous reports, but the geometrical parameters they used could be optimised further to reach much deeper colours. In this paper, we show how reinforcement learning[12-14], which belongs to artificial intelligence family, can help to find the best geometrical parameters for reaching the deepest possible red, green and blue colours for the given structure.

We see colours from objects due to their specific reflection spectra[4, 15]. This fact is used in structural colour printing to design periodic, metallic[2, 5, 16] or dielectric[17-19] nanostructures of specific dimensions to produce a desired reflection spectrum, and therefore colour. The resonances from the nanostructure can be easily controlled by changes in the geometrical properties, such as shape, size and height. Due to fabrication limitations, so far only simple shapes such as circular rods, square rods and crosses have been utilised.

Recently, neural networks have been used in the design and inverse design of nano-photonic structures[20-23], and for the design of chiral metamaterials[24]. Two methods are usually used, simple neural networks consisting of hidden layers and generative adversarial networks (GANs). Both of these methods have an undesirable property of only being able find parameters inside the limits of the data that they have been trained with, and they are unable to generate new points outside the extremities of this data. To find the extreme limits of a specific design, reinforcement learning method should be used.

Reinforcement learning, which belongs to the artificial intelligence family, is a method in which an agent tries to find the best decision at each step, based on the given rewards. Reinforcement learning has been shown to be capable of beating many Atari games[12, 25-27] and even beating human masterminds at games such as Go, backgammon and chess[28-31]. In these games, there are millions of different possible states that the agent could find themselves in, and the program should find the best action to take at each step, to maximise the chance of gaining the reward in the end. There are many branches of reinforcement learning like swarm intelligence and genetic algorithms but a human-level intelligence was achieved recently by deep Q-learnig[12]. In this paper, we use deep Q-learning to find the best parameters for designing Si based, all dielectric colour printing Fig 1(a). This model can easily be extended to other optical problems.

## Methods

In order to apply reinforcement learning to a physical problem an environment should be created that takes an action from the agent and gives a reward based on that action. Based on the reward, a neural network decides the best action at each step and keeps updating itself to continuously keep training over different states. A schematic of the algorithm is shown in Fig 1(b). The actions are defined as changing the geometrical properties of the structure, and the reward is defined based on the colour difference between the target colour and the generated colour by the structure.

### Deep Q-learning Model

The deep Q-learning model can learn the best way to act in any given situation. This method has had huge success in completing many Atari games without having to change the algorithm. This is a huge leap towards a general artificial intelligence model that can solve any given problem, like a human brain[12].

This algorithm can be explained best by using the example of playing computer games. The game has an environment that changes, depending on the actions, i.e. the input from a game controller, given to it. For each action, there is a change of state in the game. At each point, the game gives feedback though a reward system, such as, gaining points or losing a life. To write an algorithm to beat a game, we should consider all different actions at all possible states, and their relative rewards. The rule for how the actions are chosen is called a policy. The set of states, actions, and policies, form a Markov decision process, as shown below[13]:

$$s_0, a_0, r_1, s_1, a_1, r_2, ..., s_{n-1}, a_{n-1}, r_n, s_n \qquad (1)$$

where, $s$ is the state, $a$ is the action and $r$ is the reward. $S_n$ is the terminating state, which will be the target, in terms of the computer game example this would be the end state of the game. At state $s_0$, the total reward from taking specific actions can be given by the following expression, known as the discounted future reward:

$$R_t = r_t + \gamma r_{t+1} + \gamma^2 r_{t+2} + ... + \gamma^{n-t} r_n = r_t + \gamma(r_{t+1} + \gamma(r_{t+2} + ...)) = r_t + \gamma R_{t+1} \qquad (2)$$

where γ is the discount factor, a value between 0 and 1 which can be optimised as needed. The algorithm should maximise the discounted future reward ($R_t$). In deep Q-learning, this is achieved by defining a function, Q (s, a), which represents the maximum discounted future reward for each action at a given state. At each state, the action with the highest Q is chosen, this is called the policy. This function is optimised by a neural network at each step, to find the policy that can choose an action that achieves the highest future reward, even for an unknown state. The policy is defined as:

$$\pi(s) = \arg\max_a Q(s,a) \qquad (3)$$

Argmax$_a$ chooses the action for which Q is maximised. Analogous to the discounted future reward (eq. 3), this Q is defined as:

$$Q(s,a) = r + \gamma \max_{a'} Q(s',a') \qquad (4)$$

This is known as the Bellman equation[13].

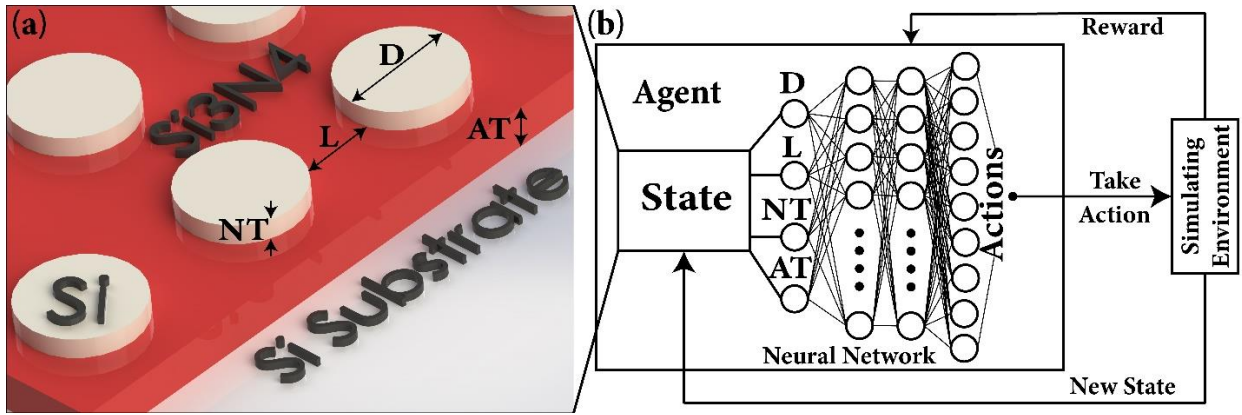

Fig. 1: Used Dielectric structure and reinforcement model. (a) A diagram of the structure used. Silicon (Si) nanodisks on an antireflective layer of $Si_3N_4$, on a silicon substrate. The diameter and thickness of nanodisks are given by D and NT, and the distance between them is given by L. The thickness of $Si_3N_4$ layer is given by AT. (b) A schematic of the design of the reinforcement model.

| Action No. | Action Definition |
|:---:|:---:|
| 0 | Decrease the spacing between disks (L) by 5nm. (min 5nm) |
| 1 | Increase the spacing between disks (L) by 5nm. (max 500nm) |
| 2 | Decrease the nanodisks diameter (D) by 5nm. (min 10nm) |
| 3 | Increase the nanodisks diameter (D) by 5nm. (max 500nm) |
| 4 | Decrease the nanodisks thickness (NT) by 5nm. (min 5nm) |
| 5 | Increase the nanodisks thickness (NT) by 5nm. (max 500nm) |
| 6 | Decrease the antireflective layer thickness (AT) by 5nm. (min 10nm) |
| 7 | Increase the antireflective layer thickness (AT) by 5nm. (max 200nm) |
| 8 | Do nothing |

Table 1: Definitions of actions used in the reinforcement model.

The final step is to connect a deep neural network, which connects the states to the Q functions. Neural networks need databases to train from. First, the model attempts to create data by exploring. After this the trained model can predict new states. To create the initial database, a method known as ε-greedy exploration was used. For each iteration, a random number (between 0 and 1) is chosen, if this number is smaller than ε (which is also between 0 and 1), then a random action is performed, if it is bigger than ε, an action determined by the network is performed. At the beginning of the learning process, ε was set to a number near 1 then decreased at each step, to a non-zero minimum, which always allows the model some chance to explore. Using this method, a database of random states and actions from which the model can be trained was built. All of the states, actions and rewards are added to memory, from which the model picks states at random to be trained on. This assures that the model remembers what it has done before and that the prediction is unbiased. A flowchart for this model is shown in Fig 1(b).

Here, we used an improved version of deep Q-learning known as double deep Q-learning[32]. This version is shown to have a better performance over normal deep Q-learning and also less overestimation in many cases. To apply this method two similar models are created instead of one main model. One of them trains the targets known as the target model and one acts as the main model as shown in Fig. 2. In each iteration, the weights of the target model is gained from the combination of the main model and the target model weights, by following formula:

$$Tw = Mw \times \tau + Tw \times (1-\tau) \tag{5}$$

Where Tw is target model weights and Mw is main model weights and $\tau$ is a hyper parameter that should be tuned by model's performance. The target which is the prediction of the model for each action (the action with the highest target will be chosen) is obtained from the addition of the targets predicted from the old state (from previous step) and the targets predicted from the new state (from current step, known as Q values) in each iteration. The main model is then trained on the current state and the target as is shown in Fig. 2.

To apply these techniques to optical problems, the following steps should be taken:

1. Define an environment which takes an action and gives a reward.
2. Define the actions.
3. Define a reward system.

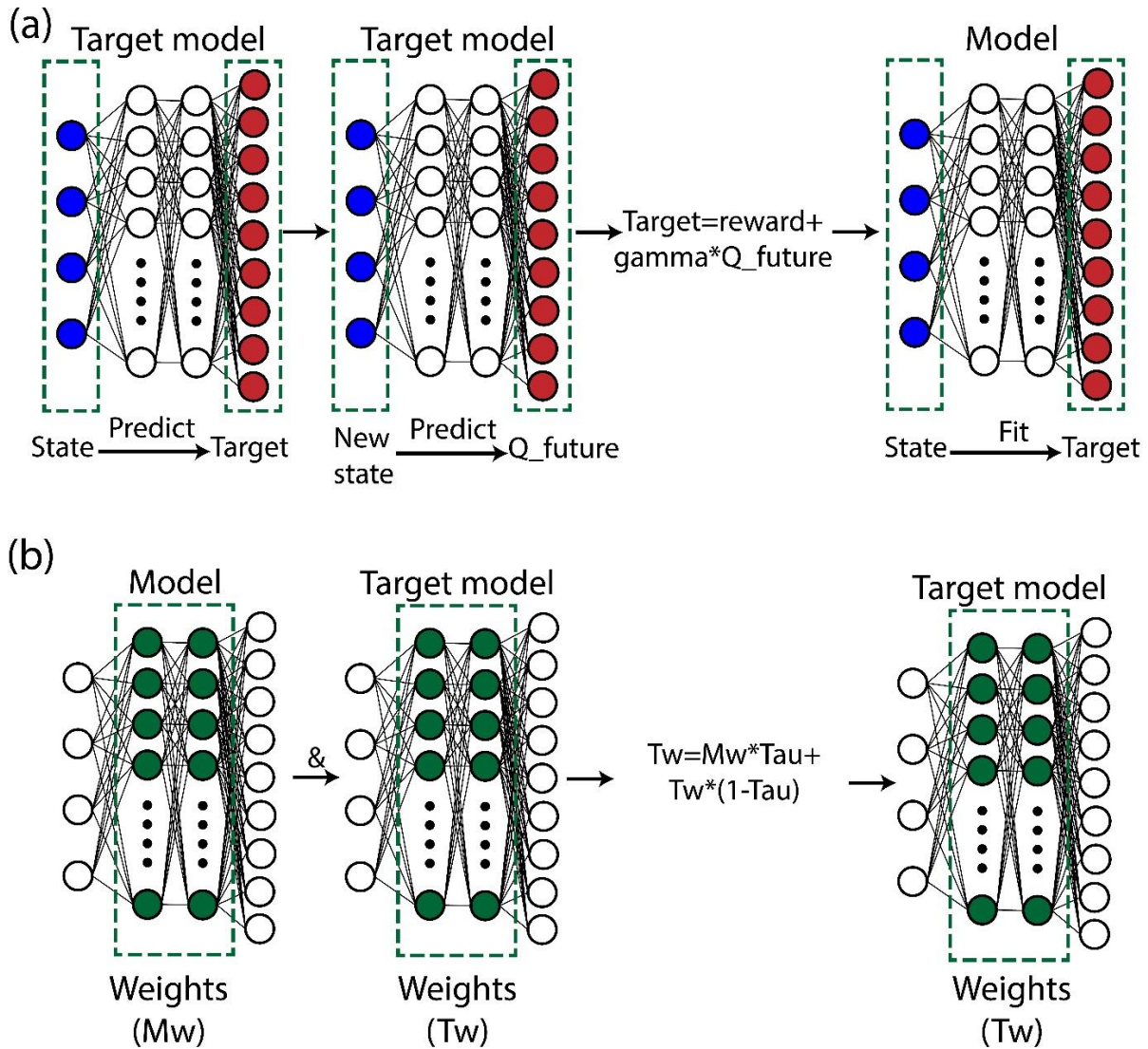

Fig. 2: The double deep Q-learning model structure. Two similar models are created. One for predicting targets named as target model and one as the main model. (a) The targets are found by the combination of the reward and the Q-values predicted by the target model. The targets found by the target model are then used to train the main model. This procedure is repeated at each iteration. (b) The weights from the main model and the target model are combined together to update the target model weights in each iteration.

**The environment and actions**

The structure is shown in Fig 1(a). This structure was proposed in recently published work[8]. Here, we show that the structure can produced much purer red, green and blue than what was reported in the paper, by further optimising the geometrical properties. The structure consists of silicon nanodisks on an antireflective silicon nitride ($Si_3N_4$) film on a silicon substrate. In the original paper, the thickness of the antireflective layer and nanodisks were kept constant, while the diameter of the nanodisks was varied from 40 to 270 nm and the spacing between the disks from 10 to 120 nm. Here, all geometrical parameters were variable and only limited by minimum and maximum sizes over a large range. The geometrical variables and limits were as follows:

- Spacing between disks (L): 5 – 500 nm; step size: 5 nm; number of steps: 99.
- Nanodisks diameter (D): 10 – 500 nm; step size: 5 nm; number of steps: 98.
- Nanodisks thickness (NT): 5 – 500 nm; step size: 5 nm; number of steps: 99.
- Antireflective layer thickness (AT): 10 – 200 nm, step size: 5 nm; number of steps: 38.

Giving a total number of states = 99 × 98 × 99 × 38 = 36,498,924. Simulating this number of possible states would require a significant amount of time and computational effort.

In order to create an environment for reinforcement learning, actions and rewards must be defined. Using the variables and limits set out above, 9 actions were defined, as shown in Table 1.

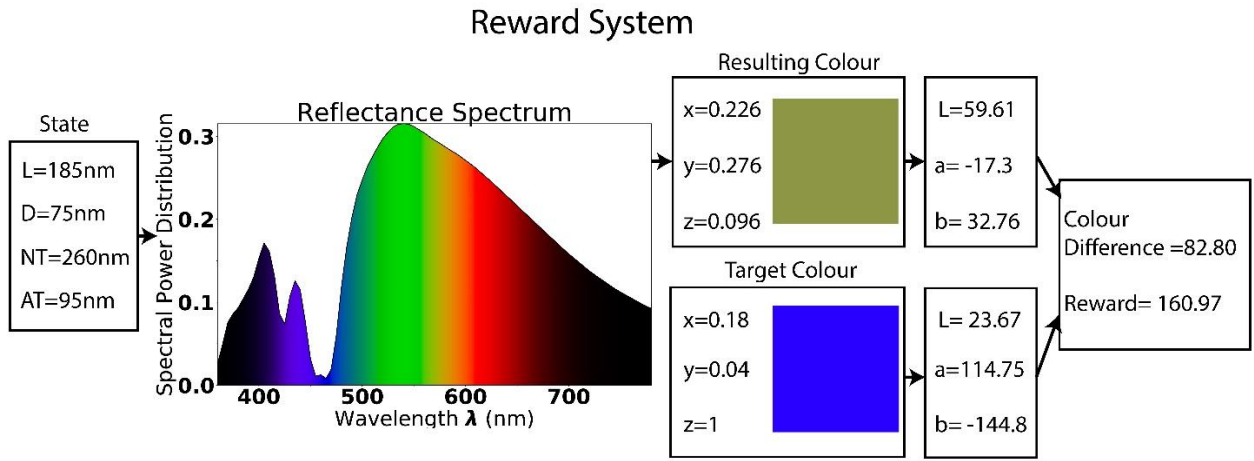

Fig. 3: The reward system used for the reinforcement model. Each state is converted to a reflectance spectrum by simulation software. The reflected spectrum is converted to XYZ colour values. The XYZ colour values are then converted to L*a*b* colour values. The reward

is calculated from the colour difference between the resulting colour from the spectrum and the target colour.

**Reward System**

The goal of the model to find the structure that generates the closest possible colour to whatever we define as the target. The reward was defined based on the smallest colour difference in L*a*b* space, between the simulated response and red, green and blue, defined as purest red, green and blue. To calculate the colour difference, first the reflection spectrum of the structure was required. The current state, the set of L, D, NT, and AT values, were sent to the simulation software. Using these values, a simulation was done and the reflection spectrum was extracted. To convert this spectrum to XYZ tristimulus colour values, defined by CIE[33], a python package named colour-science for colour operations[34] was used. As XYZ values are not linearly comparable to their respective colour, colour difference between the target and the reflection spectrum was calculated in $L^*a^*b^*$ colour space using CIE delta E 2000(CIEDE2000)[35]. Since colour difference gets smaller as closer to the target colour, but the reward should be an increasing reward, the following formula was used to calculate the reward:

$$reward = (200 - CIEDE(Obtained\ color - target\ color))^3 / 10000 \quad (6)$$

200 is big enough to make the colour difference positive. To amplify the reward as the model gets closers to the target, the difference was raised to the power of three. Dividing by 10,000 was done to make the results easier to read. After this process, the higher reward has more value than lower reward. An example of this process is shown in Fig 3.

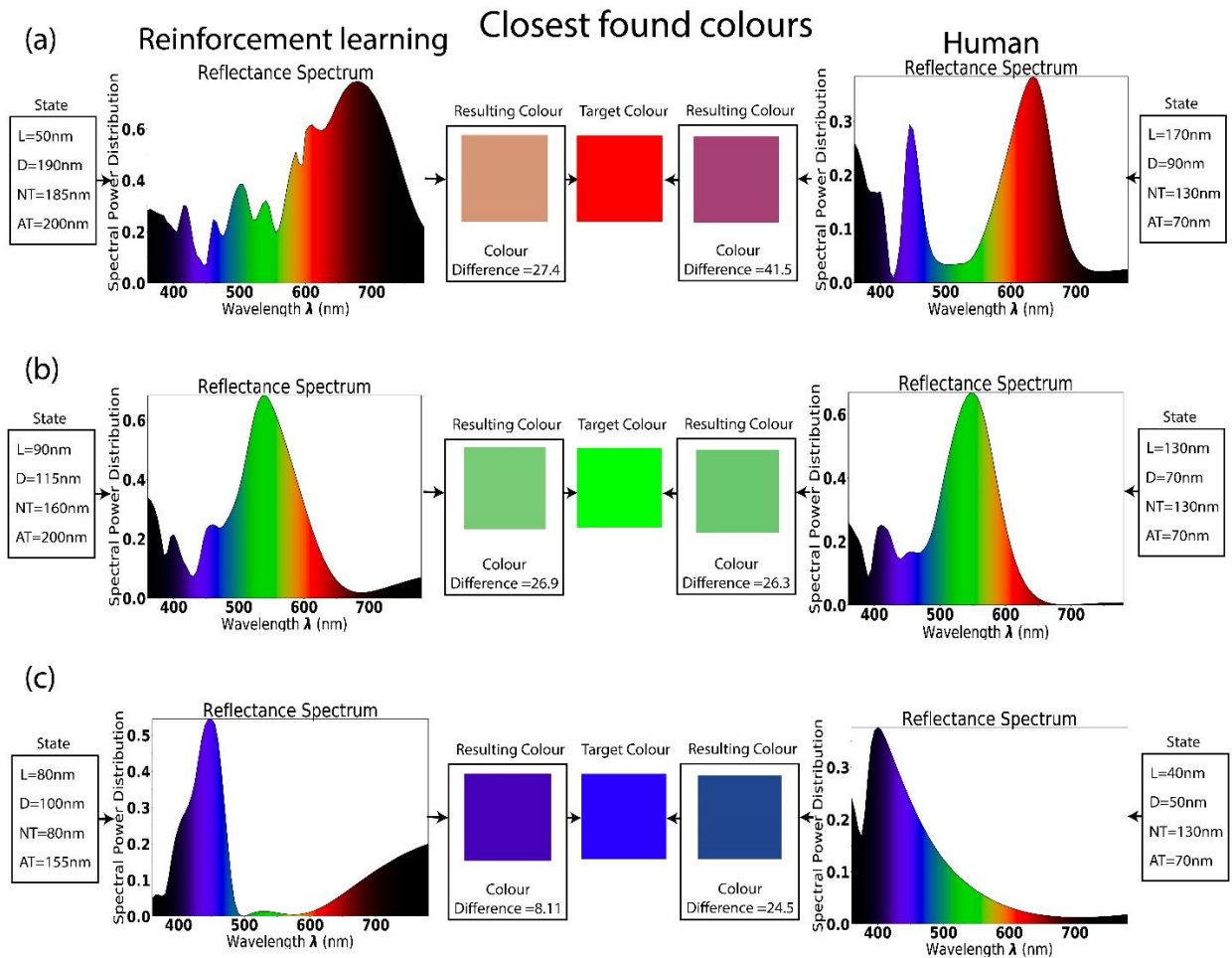

Fig. 4: The closest colours found by reinforcement learning. This figure shows the geometrical properties and the resulting colours, for red (a), green (b) and blue (c) colours found by reinforcement learning and compare them to those found by human researchers. As can be seen the RL model found much deeper colours for red and blue and almost the same green.

**Results**

All the simulations of this project were done in a commercial FDTD package, Lumerical. The code for reinforcement learning and plots are done in Python with the help of TensorFlow, Keras and colour packages. Red, green and blue colours are set as targets. As is shown in Fig. 4 the found colours by reinforcement learning in red and blue cases are much deeper than those found by humans and the green colour is almost the same. The reflection spectra and geometrical properties of the results are also shown. The best geometrical properties found by this model are L=50 nm,

D=190 nm, NT=185 nm and AT=200 nm for red, L=90 nm, D=115 nm, NT=160 nm and AT=200 nm for green, L=80 nm, D=100 nm, NT=80 nm and AT=155 nm for blue.

The whole training was done over 18 attempts containing 500 steps each, in total 9,000 steps for each of the colours. It took around one week to run the model each time, and almost all of this time was used on the simulations at each step.

A drawback of this method is that it only works for one target at a time. So at the moment it is not possible to find the best structures for red, green and blue colours simultaneously. As with other machine learning methods there are many hyper parameters to tune to achieve the best model, so finding a working model takes a long time, but once a successful model is discovered, it could be easily extended to work for similar problems.

**Discussion**

Deep Q-learning was used to find the best parameter to design structures for a physical problem, namely all dielectric, reflective colour filters. The model was able to find much more optimised dimensions for designing the structures to achieve deeper red, green and blue colours compared to the ones achieved in previous work. The same procedure could easily be extended to other physical problems and used as a tool for optimising nanosurface and nanostructure designs.

**Acknowledgments**

**Funding**

This work is financially supported by the National Research Foundation grants (NRF-2017R1E1A1A03070501, NRF-2018M3D1A1058998, NRF-2015R1A5A1037668 & CAMM-2014M3A6B3063708) funded by the Ministry of Science and ICT, Republic of Korea.

**Author Contributions**

I.S. and J.R. conceived the concept and initiated the project. I.S. designed the research and performed numerical calculations. I.S., J.R. and T.B. wrote the manuscript. All authors contributed to the discussion, analysis and confirmed the final manuscript. J.R. guided the entire project.

**Competing interests:** The authors declare that they have no competing interest.

**References**

1. Gu, Y.; Zhang, L.; Yang, J. K.; Yeo, S. P.; Qiu, C.-W., Color generation via subwavelength plasmonic nanostructures. *Nanoscale* **2015,** *7* (15), 6409-6419.
2. Tan, S. J.; Zhang, L.; Zhu, D.; Goh, X. M.; Wang, Y. M.; Kumar, K.; Qiu, C.-W.; Yang, J. K., Plasmonic color palettes for photorealistic printing with aluminum nanostructures. *Nano Letters* **2014,** *14* (7), 4023-4029.
3. Yokogawa, S.; Burgos, S. P.; Atwater, H. A., Plasmonic color filters for CMOS image sensor applications. *Nano Letters* **2012,** *12* (8), 4349-4354.
4. Kristensen, A.; Yang, J. K.; Bozhevolnyi, S. I.; Link, S.; Nordlander, P.; Halas, N. J.; Mortensen, N. A., Plasmonic colour generation. *Nature Reviews Materials* **2017,** *2* (1), 16088.

5.	Goh, X. M.; Zheng, Y.; Tan, S. J.; Zhang, L.; Kumar, K.; Qiu, C.-W.; Yang, J. K., Three-dimensional plasmonic stereoscopic prints in full colour. *Nature Communications* **2014,** *5*, 5361.
6.	Zhu, X.; Vannahme, C.; Højlund-Nielsen, E.; Mortensen, N. A.; Kristensen, A., Plasmonic colour laser printing. *Nature Nanotechnology* **2016,** *11* (4), 325-329.
7.	Roberts, A. S.; Pors, A.; Albrektsen, O.; Bozhevolnyi, S. I., Subwavelength plasmonic color printing protected for ambient use. *Nano Letters* **2014,** *14* (2), 783-787.
8.	Dong, Z.; Ho, J.; Yu, Y. F.; Fu, Y. H.; Paniagua-Dominguez, R.; Wang, S.; Kuznetsov, A. I.; Yang, J. K., Printing beyond sRGB color gamut by mimicking silicon nanostructures in free-space. *Nano letters* **2017,** *17* (12), 7620-7628.
9.	Wiecha, P. R.; Arbouet, A.; Girard, C.; Lecestre, A.; Larrieu, G.; Paillard, V., Evolutionary multi-objective optimization of colour pixels based on dielectric nanoantennas. *Nature Nanotechnology* **2017,** *12* (2), 163.
10.	Wang, L.; Ng, R. J. H.; Safari Dinachali, S.; Jalali, M.; Yu, Y.; Yang, J. K., Large area plasmonic color palettes with expanded gamut using colloidal self-assembly. *ACS Photonics* **2016,** *3* (4), 627-633.
11.	James, T. D.; Mulvaney, P.; Roberts, A., The plasmonic pixel: large area, wide gamut color reproduction using aluminum nanostructures. *Nano Letters* **2016,** *16* (6), 3817-3823.
12.	Mnih, V.; Kavukcuoglu, K.; Silver, D.; Rusu, A. A.; Veness, J.; Bellemare, M. G.; Graves, A.; Riedmiller, M.; Fidjeland, A. K.; Ostrovski, G., Human-level control through deep reinforcement learning. *Nature* **2015,** *518* (7540), 529.
13.	Sutton, R. S.; Barto, A. G., *Reinforcement learning: An introduction*. MIT press: 1998.
14.	Kaelbling, L. P.; Littman, M. L.; Moore, A. W., Reinforcement learning: A survey. *Journal of Artificial Intelligence Research* **1996,** *4*, 237-285.
15.	Busch, K.; Von Freymann, G.; Linden, S.; Mingaleev, S.; Tkeshelashvili, L.; Wegener, M., Periodic nanostructures for photonics. *Physics Reports* **2007,** *444* (3-6), 101-202.
16.	Kumar, K.; Duan, H.; Hegde, R. S.; Koh, S. C.; Wei, J. N.; Yang, J. K., Printing colour at the optical diffraction limit. *Nature Nanotechnology* **2012,** *7* (9), 557.
17.	Zhu, X.; Yan, W.; Levy, U.; Mortensen, N. A.; Kristensen, A., Resonant laser printing of structural colors on high-index dielectric metasurfaces. *Science Advances* **2017,** *3* (5), e1602487.
18.	Shen, Y.; Rinnerbauer, V.; Wang, I.; Stelmakh, V.; Joannopoulos, J. D.; Soljacic, M., Structural colors from Fano resonances. *Acs Photonics* **2015,** *2* (1), 27-32.
19.	Kim, H.; Ge, J.; Kim, J.; Choi, S.-e.; Lee, H.; Lee, H.; Park, W.; Yin, Y.; Kwon, S., Structural colour printing using a magnetically tunable and lithographically fixable photonic crystal. *Nature Photonics* **2009,** *3* (9), 534.
20.	Liu, D.; Tan, Y.; Khoram, E.; Yu, Z., Training deep neural networks for the inverse design of nanophotonic structures. *ACS Photonics* **2018,** *5* (4), 1365-1369.
21.	Peurifoy, J. E.; Shen, Y.; Jing, L.; Cano-Renteria, F.; Yang, Y.; Joannopoulos, J. D.; Tegmark, M.; Soljacic, M. In *Nanophotonic inverse design using artificial neural network*, Frontiers in Optics, Optical Society of America: 2017; p FTh4A. 4.
22.	Malkiel, I.; Nagler, A.; Mrejen, M.; Arieli, U.; Wolf, L.; Suchowski, H., Deep learning for design and retrieval of nano-photonic structures. *arXiv preprint arXiv:1702.07949* **2017**.
23.	Peurifoy, J.; Shen, Y.; Jing, L.; Yang, Y.; Cano-Renteria, F.; DeLacy, B. G.; Joannopoulos, J. D.; Tegmark, M.; Soljačić, M., Nanophotonic particle simulation and inverse design using artificial neural networks. *Science Advances* **2018,** *4* (6), eaar4206.
24.	Ma, W.; Cheng, F.; Liu, Y., Deep-Learning Enabled On-Demand Design of Chiral Metamaterials. *ACS Nano* **2018**.

25. Mnih, V.; Kavukcuoglu, K.; Silver, D.; Graves, A.; Antonoglou, I.; Wierstra, D.; Riedmiller, M., Playing atari with deep reinforcement learning. *arXiv preprint arXiv:1312.5602* **2013**.
26. Mnih, V.; Badia, A. P.; Mirza, M.; Graves, A.; Lillicrap, T.; Harley, T.; Silver, D.; Kavukcuoglu, K. In *Asynchronous methods for deep reinforcement learning*, International conference on machine learning, 2016; pp 1928-1937.
27. Lillicrap, T. P.; Hunt, J. J.; Pritzel, A.; Heess, N.; Erez, T.; Tassa, Y.; Silver, D.; Wierstra, D., Continuous control with deep reinforcement learning. *arXiv preprint arXiv:1509.02971* **2015**.
28. Silver, D.; Hubert, T.; Schrittwieser, J.; Antonoglou, I.; Lai, M.; Guez, A.; Lanctot, M.; Sifre, L.; Kumaran, D.; Graepel, T., Mastering chess and shogi by self-play with a general reinforcement learning algorithm. *arXiv preprint arXiv:1712.01815* **2017**.
29. Tesauro, G., Temporal difference learning and TD-Gammon. *Communications of the ACM* **1995,** *38* (3), 58-68.
30. Silver, D.; Huang, A.; Maddison, C. J.; Guez, A.; Sifre, L.; Van Den Driessche, G.; Schrittwieser, J.; Antonoglou, I.; Panneershelvam, V.; Lanctot, M., Mastering the game of Go with deep neural networks and tree search. *Nature* **2016,** *529* (7587), 484.
31. Silver, D.; Sutton, R. S.; Müller, M. In *Reinforcement Learning of Local Shape in the Game of Go*, IJCAI, 2007; pp 1053-1058.
32. Van Hasselt, H.; Guez, A.; Silver, D. In *Deep Reinforcement Learning with Double Q-Learning*, AAAI, Phoenix, AZ: 2016; p 5.
33. CIE, C., Commission internationale de l'eclairage proceedings, 1931. *Cambridge University Press Cambridge* **1932**.
34. Mansencal, T. M., Michael; Parsons, Michael; Canavan, Luke; Cooper, Sean; Shaw, Nick; Wheatley, Kevin; Crowson, Katherine; Lev, Ofek, Colour Science for Python 0.3.11. **2018**.
35. Sharma, G.; Wu, W.; Dalal, E. N., The CIEDE2000 color-difference formula: Implementation notes, supplementary test data, and mathematical observations. *Color Research & Application: Endorsed by Inter-Society Color Council, The Colour Group (Great Britain), Canadian Society for Color, Color Science Association of Japan, Dutch Society for the Study of Color, The Swedish Colour Centre Foundation, Colour Society of Australia, Centre Français de la Couleur* **2005,** *30* (1), 21-30.